\documentclass[prl,showpacs,preprintnumbers,twocolumn,amsmath,amssymb,superscriptaddress]{revtex4-1}

\usepackage{float}
\usepackage{color}
\usepackage{graphicx}
\usepackage{dcolumn}
\usepackage{bm}
\usepackage{epstopdf}

\newcommand{\beginsupplement}{%
        \setcounter{table}{0}
        \renewcommand{\thetable}{S\arabic{table}}%
        \setcounter{figure}{0}
        \renewcommand{\thefigure}{S\arabic{figure}}%
 \setcounter{equation}{0}
        \renewcommand{\theequation}{S\arabic{equation}}%
     }

\begin{document}

\title{Shaping a superconducting dome: Enhanced Cooper-pairing versus \\suppressed phase coherence in coupled aluminum nanograins }

\author{Uwe S. Pracht}
\email[email: ]{uwe.pracht@pi1.physik.uni-stuttgart.de}
\affiliation{1.\,Physikalisches Institut, Universit\"at Stuttgart, Germany}
\author{Nimrod Bachar}
\affiliation{Department of Quantum Matter Physics, University of Geneva, Switzerland}
\affiliation{Laboratory for Superconductivity and Optical Spectroscopy, Ariel University, Israel}
\affiliation{Raymond and Beverly Sackler School of Physics and Astronomy, Tel Aviv University, Israel}
\author{Lara Benfatto}
\affiliation{ISC-CNR and Department of Physics, Sapienza University of Rome, Italy}
\author{Guy Deutscher}
\affiliation{Raymond and Beverly Sackler School of Physics and Astronomy, Tel Aviv University, Israel}
\author{Eli Farber}
\affiliation{Laboratory for Superconductivity and Optical Spectroscopy, Ariel University, Israel}
\author{Martin Dressel}
\affiliation{1.\,Physikalisches Institut, Universit\"at Stuttgart, Germany}
\author{Marc Scheffler}
\affiliation{1.\,Physikalisches Institut, Universit\"at Stuttgart, Germany}

\date{\today}

\maketitle 
\textbf{Deterministic enhancement of the superconducting (SC) critical temperature $T_c$ is a long-standing goal in material science. One strategy is engineering a material at the nanometer scale such that quantum confinement strengthens the electron pairing, thus increasing the superconducting energy gap $\Delta$ \cite{parmenter68,croitoru07,kresin2006,garcia11,lindenfeld2011,mayoh14}, as was observed for individual nanoparticles \cite{bose10}. A true phase-coherent SC condensate, however, can exist only on larger scales and requires a finite phase stiffness $J$ \cite{Eme94}. In the case of coupled aluminium (Al) nanograins \cite{Abe66,Abe68,Deu73}, $T_c$ can exceed that of bulk Al by a factor of three, but despite several proposals the relevant mechanism at play is not yet understood. Here we use optical spectroscopy on granular Al to disentangle the evolution of the fundamental SC energy scales, $\Delta$ and $J$, as a function of grain coupling. Starting from well-coupled arrays, $\Delta$ grows with progressive grain decoupling, causing the increasing of $T_c$. As the grain-coupling is further suppressed, $\Delta$ saturates while $T_c$ decreases, concomitantly with a sharp decline of $J$. This crossover to a phase-driven SC transition is accompanied by an optical gap persisting above $T_c$. These findings identify granular Al as an ideal playground to test the basic mechanisms that enhance superconductivity by nano-inhomogeneity.}
 
\begin{figure}[H]
\begin{centering}
\includegraphics[scale=0.6]{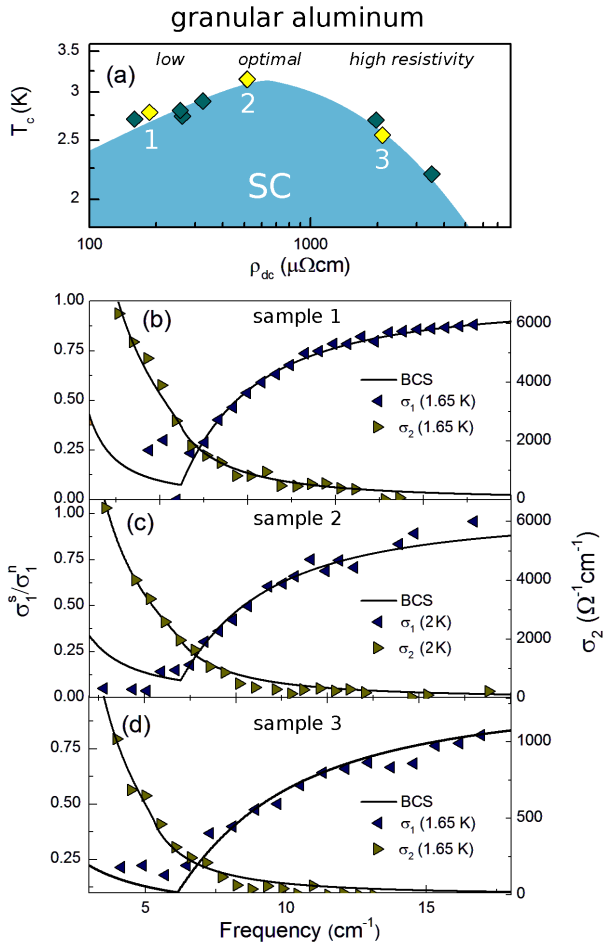}
\caption{\label{fig1} (Color online) 
\textbf{Superconducting dome and dynamical conductivity.} \textbf{(a)} Critical temperature $T_c$ as a function of the normal-state resistivity (measured at 5\,K) of granular Al films studied in this work. Yellow symbols refer to the samples displayed in panels below.  $T_c$ encloses a dome-like superconducting regime with low-, optimal- and high resistivity regimes. \textbf{(b-d)} (Normalized) spectra of $\sigma_1(\nu)$ and $\sigma_2(\nu)$ of samples located on the left (sample 1), the right (sample 3), and at the maximum (sample 2) of the SC dome. The solid lines are fits to the Mattis-Bardeen theory. Note that the fit on $\sigma_1$ disregards the low-frequency range due to excessive conductivity beyond Mattis-Bardeen theory.}
\end{centering}
\end{figure}

Bulk samples of pure Al represent a prototypical BCS superconductor (SC) with relatively low $T_{c0}\approx1.2$\,K. Several studies since the late 1960s \cite{Abe66,Abe68,Deu73} have shown a quite different situation for granular Al, i.e. thin films composed of 2\,nm grains separated by thin insulating barriers, where a superconducting condensate is established via Josephson-coupling across the grain array. The coupling between the grains can be controlled during film growth, leading to samples with strong coupling and low resistivity (LR) in electrical transport compared to high resistivity (HR) samples with weak intergrain coupling. In LR samples $T_c$ can be \emph{enhanced} up to several times  $T_{c0}$, whereas it is suppressed to zero in HR samples, shaping a superconducting dome in the phase diagram, see Fig.\,\ref{fig1}(a).\\
To understand the behavior of $T_c$ it is crucial to access the underlying SC energy scales associated with the amplitude and phase of the complex order parameter $\psi=\Delta e^{\mathrm{i}\phi}$. Indeed, while the SC energy gap $\Delta$ measures the pairing strength between the electrons, the true superfluid behavior can only be established if the Cooper pairs acquire the same macroscopic SC phase $\phi$. The energy scale controlling the rigidity of the condensate with respect to a deformation of this collective phase-coherent state is the so-called superfluid stiffness $J$. In ordinary BCS superconductors $J$ exceeds $\Delta$ by orders of magnitudes, and the SC transition at $T_c$ is amplitude-driven. However, in the unconventional situation where $\Delta$ exceeds $J$ the transition is expected to be phase-driven, due to the loss of phase coherence at a temperature scale of order of $J$. Consequently, even though several finite-size effects have been proposed to explain the enhancement of $\Delta$ in isolated nano-grains \cite{parmenter68,croitoru07,kresin2006,garcia11,lindenfeld2011}, in agreement with recent observations in Sn nanoparticles \cite{bose10}, the behavior of the global $T_c$ in large arrays of coupled grains is more difficult to assess, as here the grain coupling affects $\Delta$ and $J$ in a competing fashion. On the one hand, in the LR regime the local pairing enhancement should be strong enough to overcome the smoothening of the local density of states in the array due to strong grain coupling \cite{mayoh14}. On the other hand as  the film resistance increases, charging effects will ultimately overcome the Josephson coupling between grains, suppressing the global phase coherence.
A direct implication of the above scenario is that HR granular Al should undergo a direct superconductor-insulator transition (SIT), analogous to the one observed in homogeneously disordered films of conventional superconductors, like e.g. NbN, TiN and InO$_x$ \cite{gantmakher10,goldman15}, or suggested to occur in unconventional cuprate superconductors, where $J$ is suppressed by the proximity to the Mott-insulating phase \cite{Eme94,lee2006}. This analogy is made more interesting by the recent observation that in homogeneously disordered films \cite{sacepe08,mondal11,sacepe11,chand2012,kamlapure13,noat13}  and cuprate superconductors \cite{fisher2007,parker2010} the SC properties become spatially inhomogeneous near the SIT. In this respect, the emergent granularity, either intrinsic or extrinsic, would constitute a general mechanism for the formation of superconductivity from an almost insulating normal state, irrespective of whether the driving microscopic mechanism is based on disorder or correlation. In order to outline these potential analogies, and to better understand the mechanisms leading to a superconducting dome in granular Al, it is crucial to assess experimentally the evolution of the characteristic SC energy scales as a function of grain coupling, which so far has been only partly explored \cite{Abe68,Abr78,Ger82,Ste08,Bac14, Dyn84}.

Thin films of 40\,nm thickness were deposited on $10\times10\times2$\,mm$^3$ MgO substrates via thermal evaporation. The degree of grain coupling was tuned by controlling the O$_2$ partial pressure during deposition and quantified by the dc-resistivity $\rho_{\mathrm{dc}}$ in the normal state at 5\,K. Using a Mach-Zehnder interferometer we measured the complex transmission $\hat{t}$ of radiation in the spectral range $3-18$\,cm$^{-1}$ transmitted through the bilayer system and through a bare reference substrate, in order to disentangle Al and MgO contributions. Optical $^4$He-cryostats allowed cooling the sample to $T\geq 1.65$\,K \cite{Pra13}. At the same time we measured the dc-resistance $R_\mathrm{dc}(T)$ (four-point geometry) to obtain $T_c$ and to exclude heating upon THz absorption. In total, we examined nine samples with different resistivity values covering both sides of the SC dome, see Fig.\,\ref{fig1}(a). We measured $\hat{t}$ of all samples in the normal state well above $T_c$ and in the SC state at  $T_\mathrm{base}=1.65$\,K, as well as the temperature dependence of $\hat{t}$ for representative HR and LR samples. The complex dynamical conductivity,  $\hat{\sigma}(\nu)=\sigma_1(\nu)+i\sigma_2(\nu)$ is calculated from the optical data using  the Fresnel functions \cite{Pra13,Dre02}. \\

The great advantage of THz spectroscopy over other techniques is that it allows to simultaneously extract the fundamental energy scales of interest, $\Delta$ and $J$, from the measured $\sigma_1(\nu)$ and $\sigma_2(\nu)$ . Indeed, while the pair-breaking energy scale fixes the threshold for optical absorption in $\sigma_1(\nu)$ below $T_c$, the superfluid stiffness $J$ is directly connected to the inductive response $\sigma_2(\nu)$. In the present samples $\sigma_1(\nu)$ and $\sigma_2(\nu)$ can be adequately described  by means of the Mattis-Bardeen (MB) equations for dirty superconductors. The accuracy of the MB fit of $\sigma_1(\nu)$ and $\sigma_2(\nu)$ is shown in Fig.\,\ref{fig1}(b-d) for representative samples covering the low- (sample 1), high- (sample 3), and optimal-resisivity (sample 2) regimes of the phase diagram. Even though the fit captures well the increase of conductivity at $\nu>2\Delta/(hc)$ (where $h$ is the Planck constant and $c$ is the speed of light), it underestimates the measured $\sigma_1(\nu)$ at low frequencies. Such an excess conductivity strongly resembles the one observed, e.g., in disordered NbN and InO films \cite{crane2007,sherman15} and in cuprate films \cite{corson2000}, and is attributed to SC collective modes \cite{corson2000,stroud2000,cea2014,Swa14,sherman15}, not included in the MB theory. In the case of granular Al, where the Josephson coupling between grains is expected to be spatially inhomogeneous, this excess conductivity may be attributed to SC phase fluctuations, made optically active by disorder \cite{stroud2000,cea2014,Swa14}. 
The excess conductivity is seen in several samples at $\nu \lesssim 2\Delta/(hc)$, irrespective of the grain coupling. A more quantitative study goes beyond the scope of this work and will be considered elsewhere. The gap value $\Delta(T)$ extracted from the fit of $\sigma_1(\nu)$ is extrapolated to $T=0$ assuming the BCS temperature dependence for $\Delta(T)/\Delta(0)$, where the ratio $\Delta(0)/k_BT_c$ (with $k_B$ the Boltzmann constant) is not constrained to the weak-coupling value. 

\begin{figure}
\begin{centering}
\includegraphics[scale=0.15]{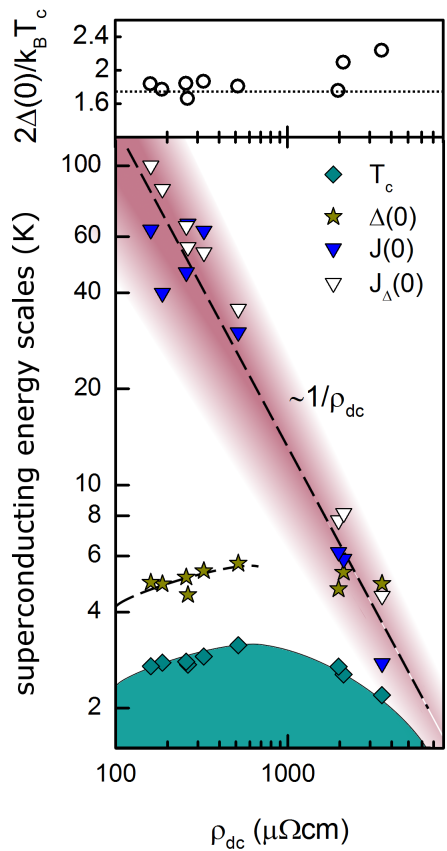}
\caption{\label{fig2} (Color online) 
\textbf{Superconducting energy scales.} $T_c,\Delta(0)$, $J(0)$, $J_\Delta(0)$ (expressed in units of temperature) and $\Delta(0)/k_BT_c$ as a function of normal-state resistivity (measured at 5\,K) of granular Al films. $T_c$ (green diamonds) encloses a superconducting dome with a maximum around $700$\,$\mu\Omega$cm where $T_c$ is enhanced by nearly a factor of three compared to the bulk value. $\Delta(0)$ (olive stars) follows the increase of $T_c$ on the left side of the dome for LR samples while it saturates in the HR regime. This is reflected in the ratio $\Delta(0)/k_BT_c$  which increases from the weak-coupling value 1.78 (dotted line) to 2.25 when crossing from the left to the right side of the dome. The calculation of superfluid stiffness from $\sigma_2(\nu)$ and from $\Delta(0)$, i.e. $J(0)$ and $J_\Delta(0)$, is subject to a uncertainty reflected by the shaded area. $J(0)$  follows approximately a  $1/\rho_\mathrm{dc}$-behavior, as expected from Eq.\,(\ref{mb}) for a constant value of $\Delta(0)$, and becomes comparable to $\Delta(0)$ in the HR regime. Dashed lines are guides to the eye.}
\end{centering}
\end{figure}

The superfluid stiffness $J$ is determined from the inductive response, i.e. $\sigma_2(\nu)$, which is proportional to $n_s/m^*$ \cite{Eme94,Supp}. More specifically, we define
\begin{equation}
\label{js}
J=\frac{\hbar^2 n_s a}{4m^*}=0.62 \times \frac{a}{\lambda^2} [K]
\end{equation}
Here, $a$ is a transverse length scale, expressed in \AA, $\lambda$ is the penetration depth in $\mu$m  and $n_s/m^*=1/\lambda^2 \mu_0 e^2$.
In an isotropic three-dimensional (3D) system, the length scale $a$ in equation\ (\ref{js}) is the SC coherence length  $\xi_0$, which is the natural cut-off for phase fluctuations, while it crosses over to the film thickness in the two-dimensional (2D) limit. Previous measurements of the upper critical field in similar granular Al samples \cite{Bac14b,Deu77} gave an estimate of $\xi_0\simeq 10$ nm,  while the analysis \cite{Deu77,Supp} of the paraconductivity above $T_c$ indicates a 2D character with an effective 2D thickness for SC fluctuations of the order of $\simeq 15$ nm throughout the phase diagram. As all these length scales are of the same order as the film thickness,  we compare the evolution of the superfluid stiffness (\ref{js}) in our samples by choosing a constant value $a=10$ nm for the sake of  simplicity. Once $n_s(T)$ is determined, the zero-temperature extrapolation follows from the two-fluid formula \cite{Supp}.

The above analysis was applied to all samples under study. Fig.\,\ref{fig2} comprises the results for the SC properties $T_c, \Delta(0),$ and $J(0)$ (all expressed in units of temperature) as well as the ratio $\Delta(0)/k_BT_c$ and presents them as functions of the normal-state resistivity. With increasing resistivity, $T_c$ is first elevated from 2.7\,K to a maximum $T_c$ of 3.15\,K before it is suppressed to 2.2\,K. This SC dome with a maximum at about $700$\,$\mu\Omega$cm is in good agreement with previous works on granular Al composed of 2\,nm grains \cite{Bac15,Bac14b,Bac13}. The enhancement of $T_c$ in the LR regime is accompanied by a concomitant {\em increase} of $\Delta(0)$, so that the ratio $\Delta(0)/k_BT_c$ remains around the weak-coupling BCS value 1.78 for all LR samples. This behavior is in sharp contrast to the usual suppression of both $T_c$ and $\Delta(0)$ for intermediate (but not too strong) disorder in homogeneously disordered films of conventional superconductors \cite{sacepe08,mondal11,sacepe11,chand2012,kamlapure13,noat13}. Thus, our measurements provide the first experimental confirmation that in granular Al the $T_c$ enhancement is due to an increase of the local pairing scale $\Delta$ in each grain by finite-size effects, as suggested by several theoretical works in the past \cite{parmenter68,croitoru07,kresin2006,garcia11,lindenfeld2011,mayoh14}. However, as the dc-resistance increases further, phase fluctuations become more prominent and the overall $T_c$ of the nanoparticle array is suppressed, even though a large local pairing survives. This is demonstrated in the upper panel of Fig. 2, where we show that on the right side of the SC dome $\Delta(0)/k_BT_c$ increases up to around 2.25, i.e., $T_c$ is reduced more strongly than $\Delta(0)$, which remains fairly constant with increasing resistance.

\begin{figure}
\includegraphics[scale=0.60]{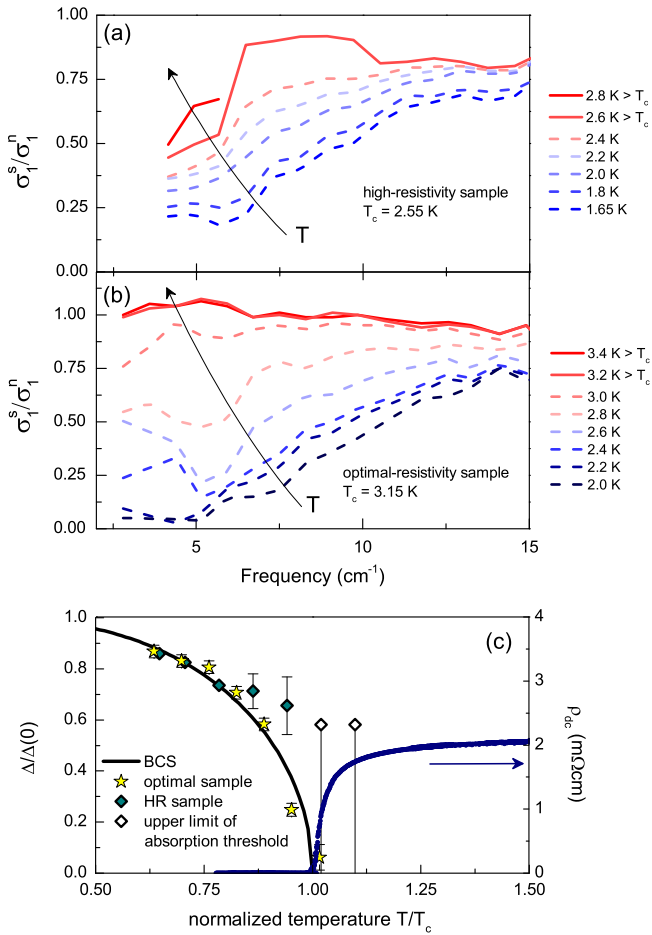}
\caption{\label{sig12_PG} (Color online) 
\textbf{Temperature evolution of spectral gap.} \textbf{(a-b)} Temperature dependence of normalized  $\sigma_1(\nu)$  of a granular Al sample in the high- and optimal resistivity regimes (sample 2 and 3 in Fig.\,\ref{fig1}(d). In case of the HR sample, the suppression of $\sigma_1(\nu)$  below $T_c=2.55$\,K (dashed lines) persists up to $T=2.8$\,K (solid lines), whereas the spectral gap closes right at $T_c$ in the LR regime. \textbf{(c)} Temperature dependence of the spectral gap for samples from the optimal (stars, sample 2) and high resistivity regimes (diamonds, sample 3). The blue data traces $\rho_{dc}(T)$ of the HR sample. For the HR sample, deviations from the BCS prediction for $\Delta(T)/\Delta(0)$ (black solid line) appear already at $T/ T_c\lesssim 1$, where $\Delta$ is anomalously large. The persistence of a gap across $T_c$ (empty diamonds) is in striking resemblance with strongly disordered or correlated superconductors.}
\end{figure}
The crossover to a SC transition driven by the loss of phase coherence in the HR samples is confirmed by the strong suppression of $J(0)$, shown in Fig.\,\ref{fig2}. The approximate scaling of $J(0)$ with $\rho^{-1}_\mathrm{dc}$ can be understood within the MB framework, where $J(0)$ can be estimated from the $\Delta(0)$ obtained from $\sigma_1(\nu)$ and the normal-state resistivity as 
\begin{equation}
\label{mb}
J_{\Delta}(0)=\frac{R_\mathrm{c}}{R_\mathrm{sq}}\frac{\pi\Delta(0)}{4},
\end{equation}
where $R_\mathrm{c}=\hbar/e^2$ and $R_\mathrm{sq}=\rho_\mathrm{dc}/a$ with same scale $a$ as used in equation (\ref{js}). Using $\Delta(0)$ extracted from $\sigma_1$ we calculate a stiffness $J_\Delta(0)$ that is similar to $J(0)$ extracted  from $\sigma_2(\nu)$ following equation\,(\ref{js}), see Fig.\,\ref{fig2}. As the grain are progressively decoupled the  $1/\rho_\mathrm{dc}$ prefactor in equation\ (\ref{mb}) varies strongly, dominating the overall scaling of $J(0)$. As a consequence, even in the LR regime  the absolute value of $J(0)$ is considerably lower than what is expected for conventional clean superconductors, where $J(0)$ scales with the Fermi energy. According to equation\ (\ref{js}) this translates to a penetration depth for LR samples of  $\lambda\approx1$\,$\mu$m \cite{Supp} much larger than $\lambda\approx 50$\,nm reported for bulk Al \cite{Bio59}. However, while in LR samples $J(0)$ still exceeds $\Delta(0)$ by one order of magnitude, in HR samples $J(0)$ is strongly suppressed and becomes comparable to $T_c$. We notice that the given estimate of $J(0)$ should be taken as an upper bound, since it neglects the additional reduction due to inhomogeneous phase fluctuations \cite{cea2014,Swa14,mayoh14}. However, the comparison with previous SQUID measurements \cite{Abr78,Ger82} of the penetration depth suggests that this effect is still quantitatively small for the samples under consideration. The slightly different values of $J(0)$ and $J_\Delta(0)$ is reflected in the shaded area in Fig.\,\ref{fig2}.\\ 

Apart from the anomalously large value of  $\Delta(0)/k_BT_c$, in the HR regime we find further evidence for an unconventional behavior of superconductivity from the dissipative conductivity. Fig.\,\ref{sig12_PG}(a) and (b) compare (normalized)  $\sigma_1(\nu)$ spectra of high- and optimal resistivity samples (sample 2 and 3 in Fig.\,\ref{fig1}) at various temperatures below (dashed lines) and above (solid lines) $T_c$. In the HR regime, MB theory agrees with the measured data at low temperatures \cite{Supp}, whereas by approaching $T_c$ the data deviate significantly from the MB prediction, with a strong suppression of  $\sigma_1(\nu)$ at low frequencies, see Fig.\,\ref{sig12_PG}(a). This suppression in $\sigma_1(\nu)$ exists up to $T=2.8$ K well above the SC transition, as evident from Fig.\,\ref{sig12_PG}(c) where $\rho_\mathrm{dc}(T)$  of the HR sample is shown for comparison. In contrast, in the case of the optimal-resistivity sample (Fig.\,\ref{sig12_PG}b) both $\sigma_1(\nu)$ and  $\hat{t}$ spectra \cite{Supp} contain no signs of a spectral gap above $T_c$, and $\Delta(T)$ follows closely the BCS temperature dependence up to $T_c$, see Fig.\,\ref{sig12_PG}c.  
For the HR sample the quality of the  MB fit degrades already at $T/T_c\simeq 0.8$, as signalled by the larger error bars in $\Delta(T)$ reported in Fig.\,\ref{sig12_PG}c, and $\Delta(T)$ evolves smoothly in a finite spectral gap found up to the highest measured temperature. The same anomalies are observed in the analysis of the paraconductivity, discussed in details in the Supplementary section \cite{Supp}. In particular, while for low- and optimal-resistivity sample the  paraconductivity is well described by ordinary Aslamazov-Larkin type of Gaussian SC fluctuations, in the HR samples we observe clear deviations that are in agreement with previous observations of an unconventional fluctuation regime based on magnetotransport and Nernst effect \cite{Bac14b}.
All these findings suggest that the anomalous $T$ dependence of $\Delta$ and $\sigma_1(\nu)$  for HR samples near $T_c$ can be attributed to a pseudogap above $T_c$, which can be viewed as a direct and natural consequence of the phase-driven transition at high resistivities, even though a full theoretical understand of it is still lacking.\\

From our comprehensive measurements of the dynamical conductivity of superconducting granular Al thin films at THz frequencies we determined the dependence of the energy scales $T_c,\Delta(0)$, and $J(0)$ on the decoupling of the Al grains. We show that decoupling promotes the individual nature of the grains and enhances the local pairing amplitude in each grain due to finite-size effects. The enhancement of both $T_c$ and $\Delta(0)$ in the low-resistivity regime is eventually overcompensated in high-resistivity samples by enhanced phase fluctuations, which suppress $T_c$ while the pairing amplitude remains large. The strong suppression of $J(0)$ and the persistence of a spectral gap above $T_c$ in the high-resistivity regime indicate a crossover to a phase-driven transition. Our results provide clear constraint on further theoretical modelling of the superconductivty in granular arrays,  that is a prerequisite to achieve full control in the $T_c$ enhancement via engineered 
inhomogeneities  in low-dimensional superconducting nanostructures.\\

We acknowledge discussions with Ya. Fominov, A. Frydman and A. Garcia-Garcia. U.S.P is grateful for financial support from the Studienstiftung des deutschen Volkes. We acknowledge financial support from the German-Israeli Foundation for Scientific Research and Development (GIF Grant no. I-1250-303.10/2014)  L.B. acknowledges financial support by Italian MIUR under projects FIRB-HybridNanoDev-RBFR1236VV,
PRIN-RIDEIRON-2012X3YFZ2, and Premiali-2012 ABNANOTECH. N.B. and U.S.P. contributed equally to this work.

\newpage{}
\beginsupplement
\onecolumngrid
\begin{centering}
\textbf{Supplementary material for: \\
Shaping a superconducting dome: Enhanced Cooper-pairing versus suppressed phase coherence in coupled aluminum nanograins\\}
\end{centering}

\subsection{Analysis of the conductivity and determination of $\Delta(0)$ and $J(0)$}

\begin{figure}[htb]
\includegraphics[scale=0.8]{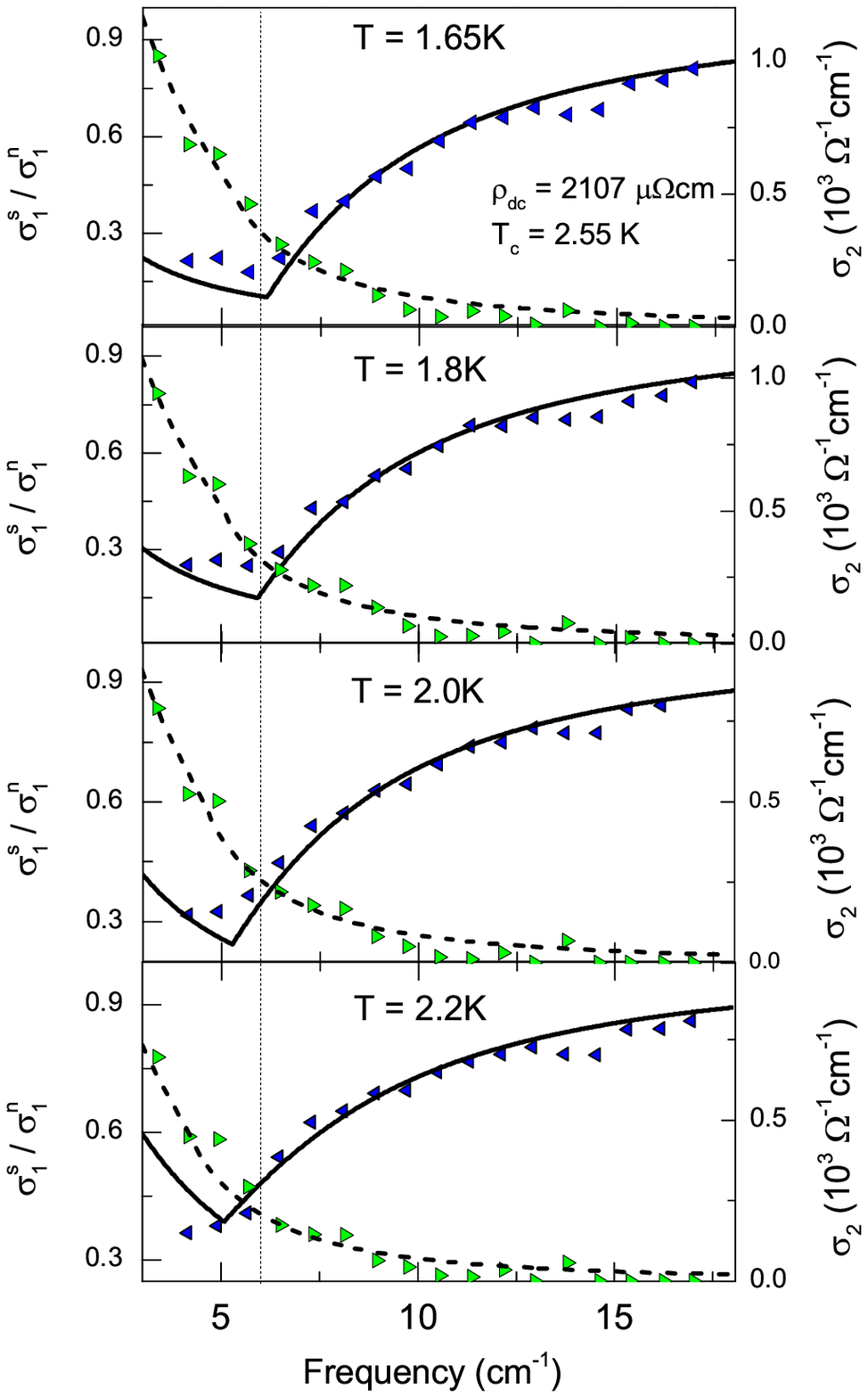}
\caption{\label{fig:sig12_SM} (Color online) 
$\sigma_1(\nu)$ (blue triangles) and $\sigma_2(\nu)$ (green triangles) of a HR sample as functions of frequency at various temperatures in the superconducting state together with fits to MB formula (\ref{s1}) and (\ref{s2}). Apart from moderate excess conductivity at subgap frequencies (denoted by the dashed line, excluded from the $\sigma_1$ fit) MB theory reproduces our results remarkably well. Corresponding spectra of an LR sample can be  found elsewhere \cite{Bac14}. }
\end{figure}

We analyze both $\sigma_1(\nu)$ and $\sigma_2(\nu)$ by means of the ordinary Mattis-Bardeen (MB) formulas \cite{mb58} for the optical conductivity:
\begin{eqnarray}
\label{s1}
\frac{\sigma_1(\nu)}{\sigma_n}&=&\frac{\pi n_s}{m^*\sigma_n}\delta(\nu)+\frac{2}{h\nu} \int_{\Delta}^\infty dE g(E)\left[ f(E)-f(E+h\nu)\right]
-\frac{\Theta(h\nu-2\Delta)}{h\nu}\int_{\Delta-h\nu}^{-\Delta} dE g(E) \left[1-2f(E+h\nu)\right]\\
\label{s2}
\frac{\sigma_2(\nu)}{\sigma_n}&=&\frac{1}{h\nu} \int_{-\Delta,\Delta-h\nu}^{\Delta} dE g(E)\left[ 1-2f(E+h\nu)\right]
\frac{E(E+h\nu)+\Delta^2}{\sqrt{\Delta^2-E^2}\sqrt{(E+h\nu)^2-\Delta^2}},
\end{eqnarray}
where the function $g(E)$ is
\begin{equation}
\label{ge}
g(E)=\frac{E(E+h\nu)+\Delta^2}{\sqrt{E^2-\Delta^2}\sqrt{(E+h\nu)^2-\Delta^2}}.
\end{equation}
Here $\sigma_n$ indicates the normal-state conductivity, that is featureless in the THz frequency range for our samples. To exclude any residual fluctuation effect we take as $\sigma_n$ the one measured at 5 K, well above $T_c$ for all samples. In Fig.\ \ref{fig:sig12_SM} we show the accuracy of the fit of $\sigma_1(\nu)$ and $\sigma_2(\nu)$ of a representative high-resistivity (HR) sample at several temperatures. We remark that the fits show deviations at frequency $\nu\lesssim 2\Delta/(hc)$, where we observe an excess conductivity analogous to the one  observed in other conventional superconductors as NbN and InO \cite{crane2007,sherman15}. This has been attributed to superconducting collective modes \cite{corson2000,stroud2000,cea2014,Swa14,sherman15}, not included in the MB theory, whose detailed study goes beyond the scope of this work and will be considered elsewhere. The obtained values of $\Delta(T)$ are fitted with a conventional BCS temperature dependence for $\Delta(T)/\Delta(0)$ vs $T/T_c$, where $\Delta(0)/k_BT_c$ enters as the only free parameter. The BCS expression reproduces accurately the behavior of $\Delta(T)$ for  samples on the left side of the superconducting dome, while for HR samples some deviations occur near $T_c$, that can be attributed pseudogap above $T_c$ in HR samples. 

\begin{figure*}
\includegraphics[scale=0.15]{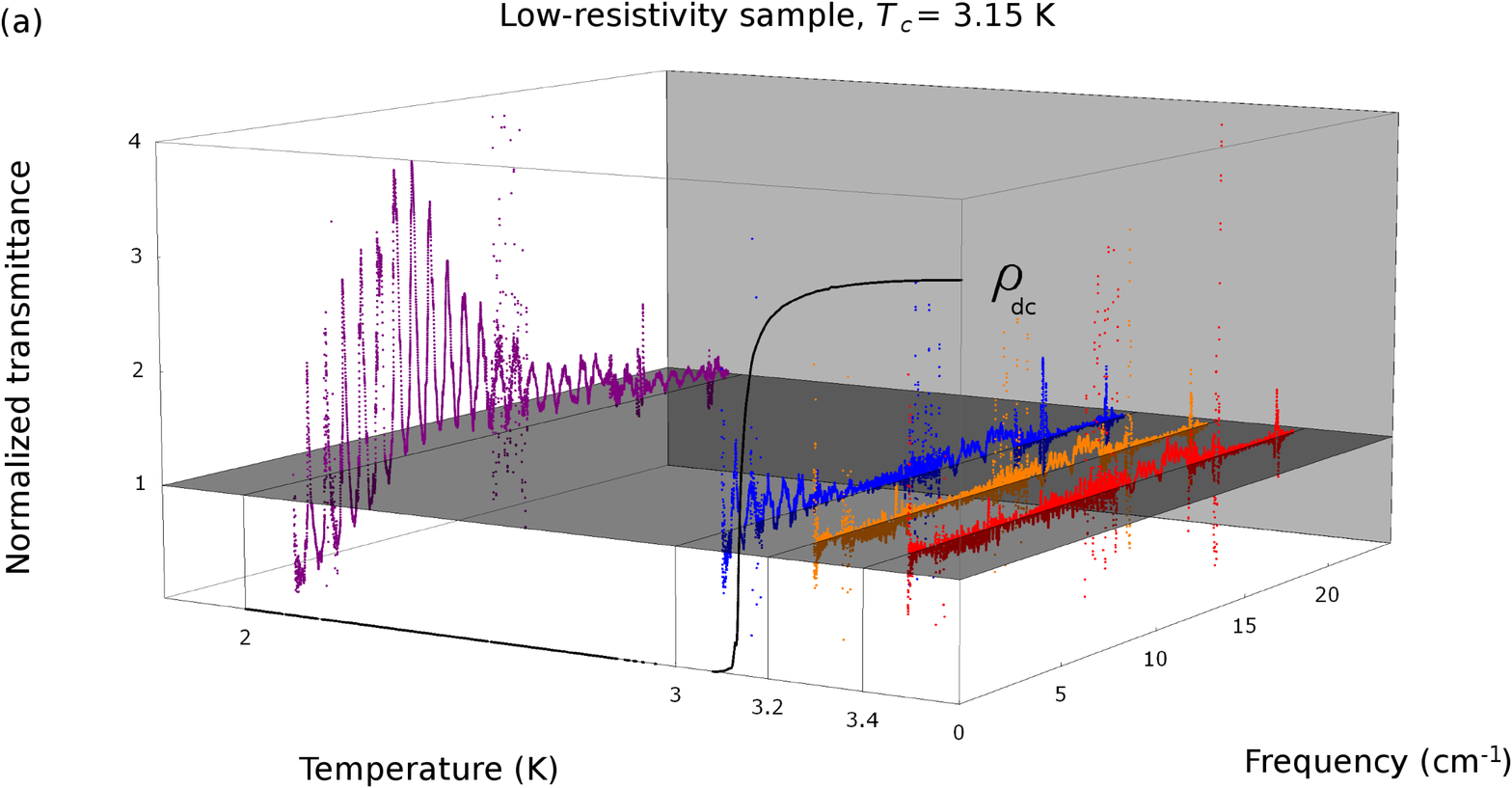}
\includegraphics[scale=0.15]{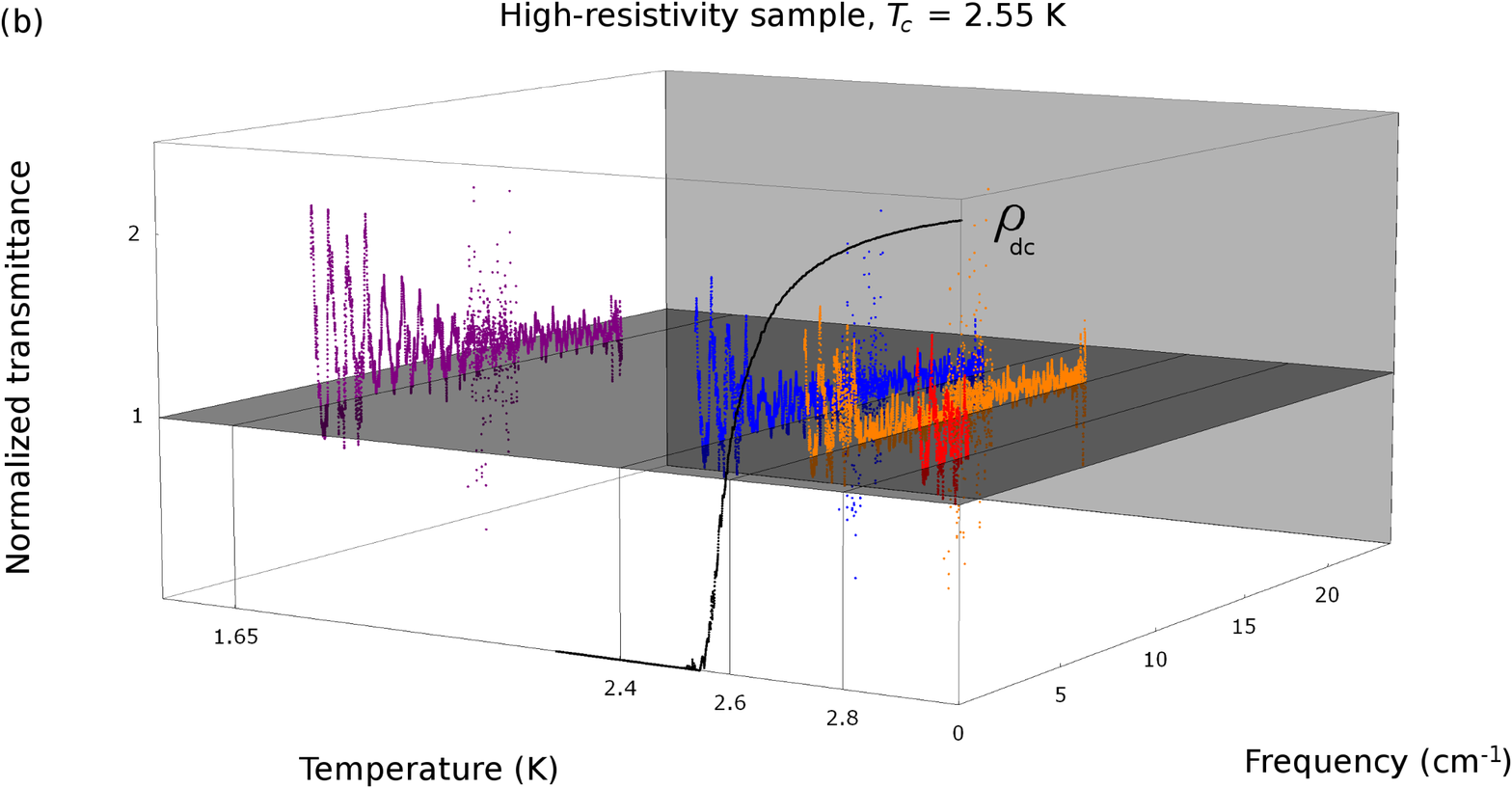}
\caption{(Color online) Normalized transmittance  $\tilde{t}=|t|/|t_0|$ of granular Al films versus frequency and selected temperatures. The thick solid line sketches the temperature dependence of the dc-sheet resistivity $\rho_{dc}$.\textbf{(a)} Low-resistivity ($T_c=3.15$\,K,$\rho_\mathrm{dc}= 516$\,$\mu\Omega$cm) and \textbf{(b)} high-resistivity sample ($T_c=2.55$\,K, $\rho_\mathrm{dc}= 2107$\,$\mu\Omega$cm). For the LR sample, we find $|t|/|t_0|$ to increase above unity only below $T_c$ in sharp contrast to the HR regime, where $|t/t_0|>1$ already above $T_c$ indicates a pseudogap.}
\end{figure*}

\begin{figure}
\includegraphics[scale=0.6]{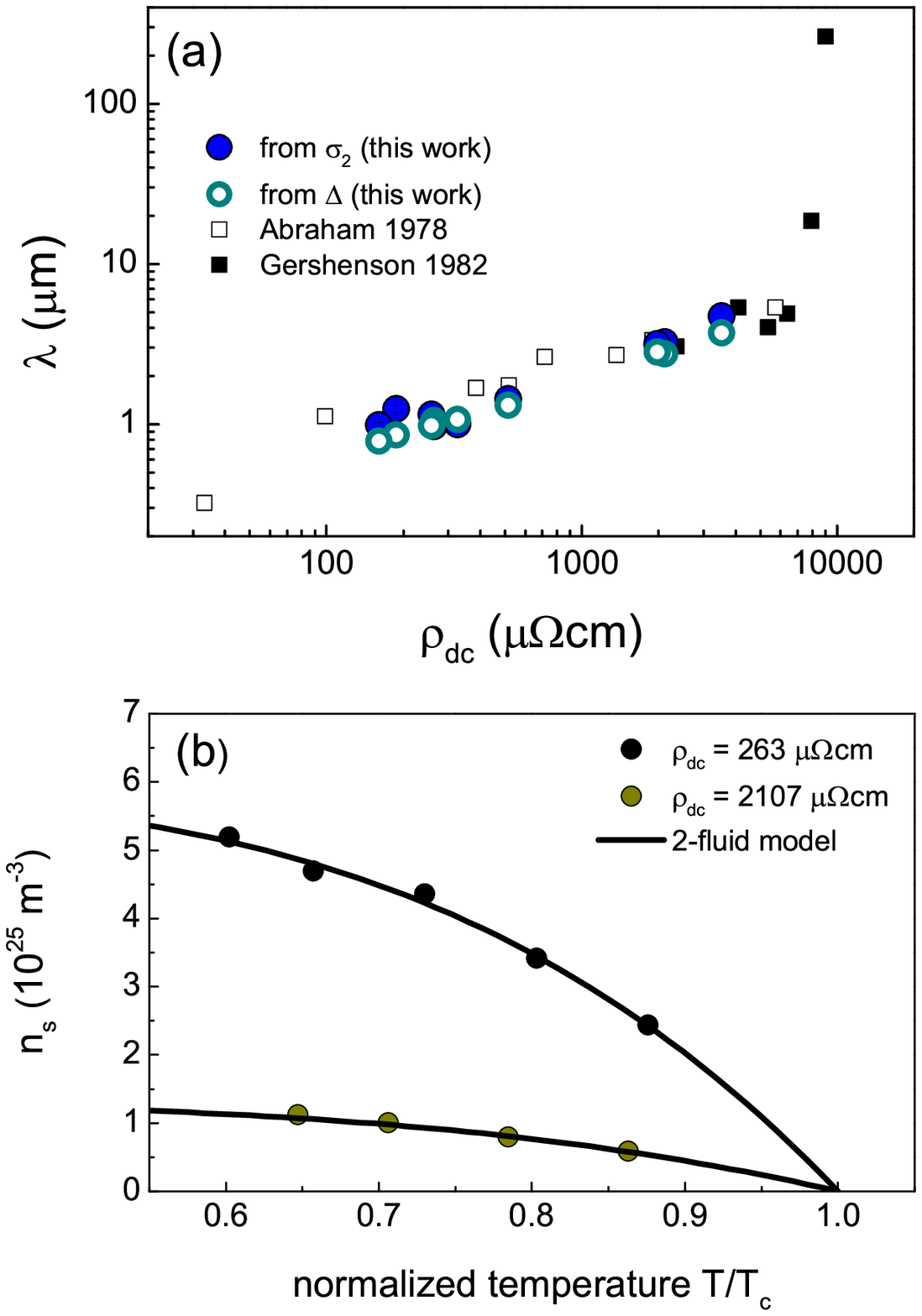}
\caption{\label{fig:lambda} (Color online) 
\textbf{(a)} Penetration depth $\lambda$ versus normal-state resistance. The present work (colored dots) very well reproduces results obtained in previous works by of Abraham\cite{Abr78} and Gershenson\cite{Ger82} (empty and filled squares), where the inverse penetration depth has been directly measured. In addition, we note that the absolute numbers of $\lambda$ obtained from the inductive response (blue dots) are nearly identical with the calculation of $\lambda$ from $\Delta$ (green dots) within MB theory. \textbf{(b)} Superfluid density $n_s$ of a LR and HR sample versus temperature. The solid lines are fits to the two-fluids approximation (\ref{eq:2fluid}) to obtain the zero-temperature extrapolation $n_{s,0}$. For the LR and HR samples we estimated a mass enhancement of $m^*/m  =2$ and 5, respectively, from the normal-state resistivity \cite{Bac15}. }
\end{figure}
As a check of consistency of this procedure we remark that the emergence of a global superconducting state can directly be seen from the raw transmittance (transmission amplitude) $|t|$ of the electromagnetic radiation passing through the Al film. Below $T_c$ and in the spectral range relevant for this work, $|t|$ is higher than the normal-state transmittance $|t_0|$ and frequency dependent with a maximum close to the spectral gap $2\Delta$, i.e. the normalized transmittance $\tilde{t}=|t|/|t_0|$ is above unity. The presence of a pseudogap above $T_c$ manifests as  $\tilde{t}=|t|/|t_0|>1$ in the low frequency limit, or, equivalently $\tilde{t}=1$ in case of no pseudogap present. In Fig. 3 we show  $\tilde{t}$ for a low-resistivity (LR) sample ($\rho_\mathrm{dc}= 516$\,$\mu\Omega$cm) on the left side of the superconducting dome for various temperatures above and below $T_c=3.15$\,K. While at low temperatures  $\tilde{t}$ is well above unity, any signature of superconducivity vanishes directly at $T_c$. This is contrasted by a HR sample, Fig. 3b, where we find $\tilde{t}$ exceeding  unity already at 2.8\,K well above $T_c=2.55$\,K which suggests the opening of a pseudogap that smoothly develops into the superconducting gap.

The superfluid density  $n_s$ is given by the missing finite-frequency spectral weight between normal- and superconducting state $\sigma_1(\nu>0)$, and it controls the weight of the zero-frequency delta response in Eq.\ (\ref{s1}). Since it cannot be directly accessed by optical probes, we extract it from the Kramers-Kronig transform $\sigma_2(\nu)$ as
\begin{equation}
n_s=\frac{2\pi m^*}{e^2} \lim_{\nu\to 0}\nu\sigma_2(\nu),
\end{equation}   
where $m^*$ and $e$ are the effective electron mass and charge. We fit the experimental data in the THz range to the MB formula (\ref{s2}), see Fig.\ \ref{fig:sig12_SM}, and extrapolate fit times frequency to zero. The value at zero temperature is then obtained from a standard 
two-fluids temperature dependence:
\begin{equation}
\frac{n_s}{n_{s,0}}\approx 1-\left(\frac{T}{T_c}\right)^4\label{eq:2fluid},
\end{equation} 
where $n_{s,0}$ is the zero-temperature superfluid density. As in the case of $\Delta(T)$, we find representative samples on both sides of the superconducting dome following this universal dependence on $T/T_c$, see Fig.\ \ref{fig:lambda}b), which justifies to calculate  $n_{s,0}$ from Eq. (\ref{eq:2fluid}) for samples where no full $T$-dependence was measured. We remark that even though the exact value of $m^*$ is needed to extract $n_s$ from the measured $\sigma_2(\nu)$, it is instead irrelevant when computing the superfluid stiffness $J(0)=\hbar^2 n_{s,0} a/4m^*$, which only depends on the ratio $n_s/m^*$.  We also notice that the separate analysis of $\sigma_2(\nu)$ to obtain $n_{s,0}$ is in principle redundant, since within the MB theory we could use directly the value of $\Delta(0)$ extracted from the $\sigma_1$ fits to determine the zero-temperature inductive response. Indeed, from Eq.\ (\ref{s2}) one immediately sees that  $\sigma_2(\nu\to 0,T=0)=\pi \Delta(0)\sigma_n/(h\nu)$. Since in the dirty limit $\sigma_n$ 
coincides with $1/\rho_{dc}$ in the THz frequency range, we can estimate $n_{s,0}$ as:
\begin{equation}
\label{mb}
n_{s,0}^\Delta=\frac{2\pi m^*}{e^2}\frac{\pi \Delta(0)}{\hbar \rho_{dc}}
\end{equation}
so that the corresponding estimate $J_\Delta$ of the stiffness corresponds to Eq. (2) of the main text. However, since the deviations of $\sigma_1(\nu)$ from the MB behavior occurs exactly below $2\Delta$, we analyzed $\sigma_2(\nu)$ independently on $\sigma_1(\nu)$, and we cross-check afterwards the consistency between the two approaches, see Fig.\ \ref{fig:lambda}a). Here the extracted values of $n_{s,0}$ and $n_{s,0}^\Delta$ are converted to the penetration depth $\lambda=\mu_0e^2n_s/m^*$ (with $\mu_0$ the vacuum permeability), in order to compare them with direct measurements of $\lambda$  done in previous works \cite{Abr78,Ger82}. Both estimates of $\lambda$ are consistent with each other and they are in very good agreement with previous findings. This shows also that the quantitative suppression of $n_{s,0}$ due to the collective-mode contribution below $2\Delta$ in $\sigma_2(\nu)$ is relatively small, and it justifies the use of the MB formula to extrapolate $\sigma_2$ to zero frequency.

\subsection{Analysis of the paraconductivity above $T_c$}

\begin{figure*}[htb]
\centering
\includegraphics[scale=0.83]{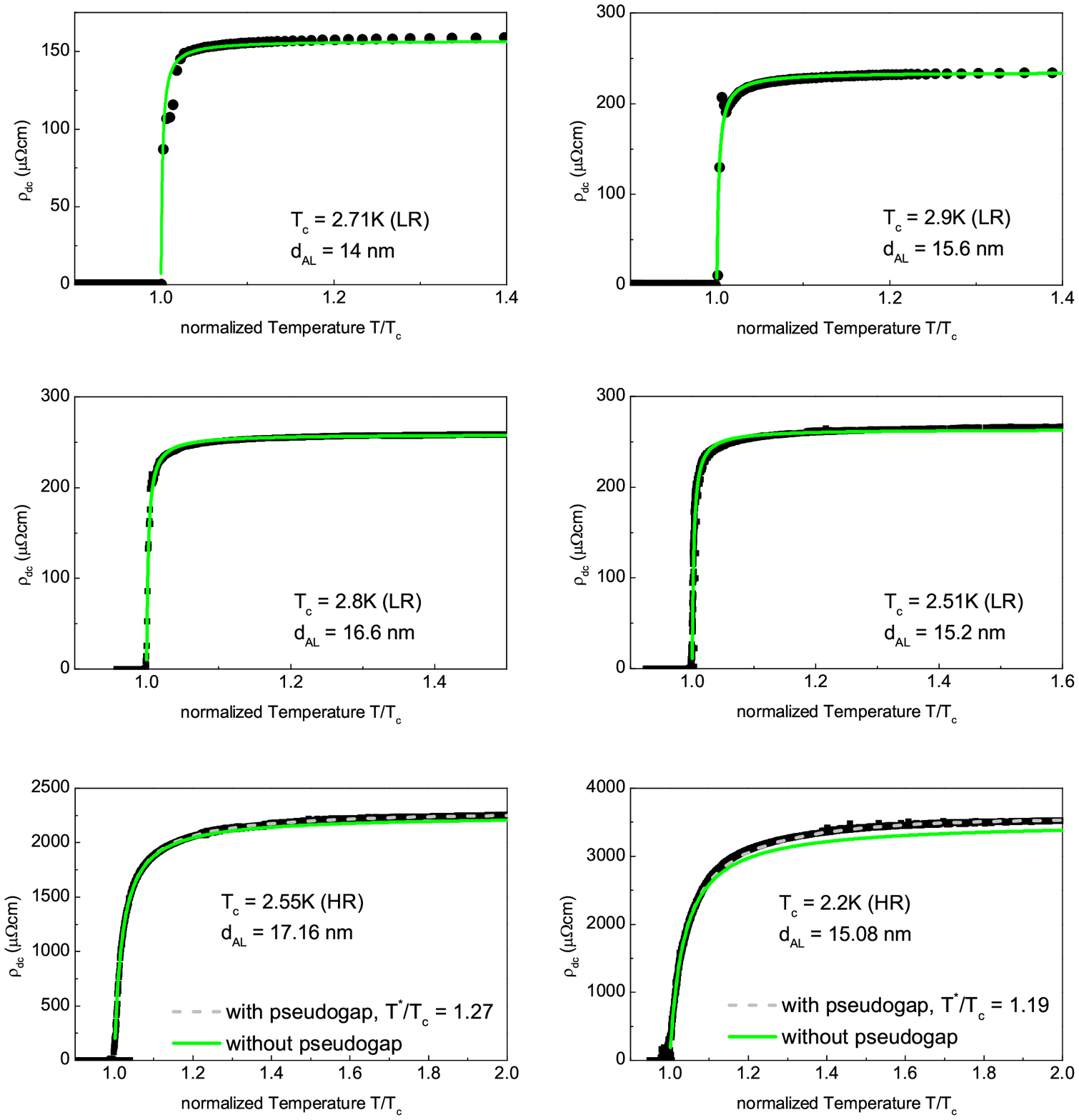}
\caption{Fit of the resistivity with the AL formula (\ref{al2d}) for LR samples and with the modified AL formula (\ref{al2dpg}) for the HR samples. The parameters of the fit are shown in each panel.}
\label{fig-alfit}
\vspace{1cm}
\end{figure*}

To analyze the effect of the superconducting fluctuations above $T_c$  (here defined as the temperature where the resistance becomes immeasurably small) we focus on the Aslamazov-Larkin (AL) contribution to the paraconductivity $\Delta\sigma$ \cite{VarlamovBook}, that is the most relevant near $T_c$. In agreement with previous work\cite{Bac14b}, we found that in our samples the paraconductivity has the temperature dependence expected in 2D, so that \cite{VarlamovBook}:
\begin{eqnarray}
\label{al2d}
\frac{\Delta\sigma_{AL}}{\sigma_N} =\frac{e^2}{16\hbar d_{AL}\sigma_N}\frac{1}{\epsilon}=\frac{R_\square}{16 R_c}\frac{1}{\epsilon}, 
\end{eqnarray}
where $R_c=\hbar/e^2$, $R_\square=\rho_{dc}/d_{AL}$ and $d_{AL}$ is a transverse length scale, of the order of the film thickness, that determines the effective  2D unit for SC fluctuations in our granular samples. The parameter $\epsilon$ contains the temperature dependence, and in the BCS limit it is given by
\begin{equation}
\label{epsbcs}
\epsilon=\ln \frac{T}{T_c}.
\end{equation}
We fitted all the data using the 2D AL formula (\ref{al2d}), using $d_{AL}$ and $T_c$ as a free parameters. As one can see in Fig.\ \ref{fig-alfit} the agreement with the data in the LR regime is remarkably good, with an effective thickness $d_{AL}\simeq 14-17$ nm as an adjustable fit parameter, that is within a factor of 2-3 from the real film thickness. Given the granular nature of the film this is a reasonable approximation, considering that the fit reproduces the data up to temperatures as large as twice $T_c$ without any other adjustable parameter. 
On the other hand, when one analyzes the HR films, two remarkable differences arise: (i) the resistivity is not completely saturated up to temperatures as large as twice $T_c$; (ii) the fit with the AL formula fails around $T\simeq 1.2 T_c$, since the experimentally measured paraconductivity decays faster than predicted by   Eq.\ (\ref{al2d}). Interestingly, the very same behavior has been observed also in underdoped cuprates for  samples in the pseudogap regime. In this case, it has been observed that a very good formula that works both near and far from $T_c$ is the following one \cite{caprara2005}:
\begin{equation}
\label{al2dpg}
\frac{\Delta\sigma_{AL}}{\sigma_N} =\frac{R_\square}{16 R_c}\frac{1}{\epsilon_0\sinh(\epsilon/\epsilon_0)}, \quad \epsilon_0=\ln\frac{T^*}{T_c},
\end{equation}
that reduces to the usual one (\ref{al2d}) when $\epsilon\ll \epsilon_0$, so that $\sinh(\epsilon/\epsilon_0)\simeq \epsilon/\epsilon_0$, but decays faster for $\epsilon\gg \epsilon_0$. As it has been discussed in the context of cuprates\cite{caprara2005,varlamov2011}, such a suppression of paraconductivity with respect to the standard formula (\ref{al2d}) can be indeed explained assuming that a pseudogap survives in the electronic Green's function up to a temperature $T^*$ larger than $T_c$. In the case of granular Al the formula (\ref{al2dpg}) works remarkably well for the two most disordered films, see Fig.\ \ref{fig-alfit}, where pseudogap signatures have been reported also from the MB analysis of the optical spectra.\\

\end{document}